# Room-Temperature Plasmon-Assisted Resonant THz Detection in Single-layer Graphene Transistors

José M. Caridad[1,2], Óscar Castelló[1,2], Sofía M. López Baptista[1], Takashi Taniguchi[3], Kenji Watanabe[4], Hartmut G. Roskos[5], Juan A. Delgado-Notario[1*]

[1] Department of Applied Physics, University of Salamanca, Salamanca, 37008, Spain

[2] Unidad de Excelencia en Luz y Materia Estructurada (LUMES), Universidad de Salamanca, Salamanca, 37008, Spain

[3] Research Center for Materials Nanoarchitectonics, National Institute for Materials Science, 1-1 Namiki, Tsukuba, 305-0044, Japan

[4] Research Center for Electronic and Optical Materials, National Institute for Materials Science, 1-1 Namiki, Tsukuba, 305-0044, Japan

[5] Physikalisches Institut, Johann Wolfgang Goethe-Universität, Max-von-Laue-Str. 1, Frankfurt am Main, D-60438, Germany

*corresponding author: juanandn@usal.es

# Abstract

Frequency-selective or even frequency-tunable Terahertz (THz) photodevices are critical components for many technological applications that require nanoscale manipulation, control and confinement of light. Within this context, gate-tunable phototransistors based on plasmonic resonances are often regarded as the most promising devices for frequency-selective detection of THz fields. The exploitation of constructive interference of plasma waves in such detectors not only promises frequency selectivity, but also a pronounced sensitivity enhancement at the target frequencies. However, clear signatures of plasmon-assisted resonances in THz detectors have been only revealed at cryogenic temperatures so far, and remain unobserved at application-relevant room-temperature conditions. In this work, we demonstrate the sought-after room-temperature resonant detection of THz radiation in short-channel gated photodetectors made from high-quality single-layer graphene. The survival of this intriguing resonant regime at room-temperature ultimately relies on the weak intrinsic electron-phonon scattering in graphene, which avoids the damping of the plasma oscillations.

# Introduction

Terahertz (THz) radiation (0.1 – 10 THz) has a strong perspective in a wide range of different applications, including metrology and characterization of nanomaterials[1], upcoming 6G wireless communications[2], non-invasive imaging[3], biosensing[4], high-resolution spectroscopy[5], together with many others[6,7]. An emerging and important research area within THz technology is the study of novel, efficient and functional photodetectors operating at these frequencies[8]. The majority of photodetectors reported up to date (if not all), including sensors made of many different nanomaterials[9–15], operate either in broadband mode (i.e. without being selective to a given frequency) at room temperature, or over narrow fixed frequency bands (i.e. without being frequency tunable), for example, by embedding antennas in the detector. Frequency-tunable THz photodetectors working at atmospheric conditions are therefore unavailable so far, despite being desirable components to *i)* boost the performance of some applications at specific and selected THz wavelengths[16] and *ii)* provide new functionalities such as selective sensing, frequency mixing, multiplication, and modulation as well as nanoscale confinement of light[17].

One of the most prominent ideas to design tunable and selective THz photodetectors, originally introduced by M. Dyakonov and M. Shur more than two decades ago, predicts that two-dimensional (2D) gated FETs may exhibit a resonant response to electromagnetic THz radiation at discrete plasma oscillation frequencies of the 2D electrons in the device channel[17]. In this pioneering proposal, the resonant operation of FET photodetectors is univocally defined by a quality factor, $Q = \omega\tau$, which must be much larger than unity ($Q = \omega\tau \gg 1$, where $\omega = 2\pi f$ with *f* being the frequency of the incoming radiation and $\tau$ the momentum relaxing scattering time of charge carriers in the system, respectively). In other words, resonant THz photodetection should arise in plasmonic FETs at any temperature when a negligible damping of the plasma waves occurs in the channel. In such conditions, the device channel acts as a tunable plasmonic cavity with a set of multiple resonances defined by the incoming frequency, the device length and the density of charge carriers in the system[17]. This exotic regime is in clear contrast to the more commonly observed and studied broadband (non-resonant or overdamped) case[18–23], characterized by a $Q \ll 1$ with plasmons being strongly damped in the channel and even decaying long before reaching the other side of the plasmonic cavity.

Up to date, several experimental studies have attempted to demonstrate resonant THz detection in different 2D electron gases systems with varying levels of success. Convincing signatures of plasmon resonances, including the appearance of frequency-dependent oscillations in the zero-bias photoresponse of the system w.r.t. the carrier density, have been identified at cryogenic temperatures in FET devices made of some high-quality semiconductors such as III-V materials[24–27] and bilayer graphene[28]. However, such features vanish rapidly when operating above cryogenic temperatures and long before reaching room-temperature. This fact notably limits the potential use of resonant THz photodetectors for real life applications[6,7].

In this letter, we demonstrate room-temperature THz detection in FET devices made of high-quality, single-layer graphene. In particular, we show how the characteristic frequency-dependent oscillations in the photoresponse of monolayer graphene FETs are largely tunable with the density of charge carriers in the device (i.e. with applied top gate voltage), unique fingerprints of the resonant detection which are furthermore visible from cryogenic up to room temperature. The fact that these robust signatures persist up to 300K can be directly ascribed to the weak acoustic phonon scattering in monolayer graphene, which leads to large carrier mobility values in the material even at elevated temperatures[29,30]. In other words, as shown below, the resonant condition $Q \gg 1$ is also fulfilled in high-quality, single-layer graphene FET detectors at room-temperature.

## Experimental Details

In order to observe plasmonic resonant THz detection, we fabricated a short-channel (length $L_{ch}$ = 6 µm) dual-gate, high-mobility, single-layer graphene FET device (Figure 1 (a)) by using a state-of-the-art dry-stacking technique[22] to encapsulate a mechanically exfoliated single-layer graphene sheet (hereafter referred to as graphene) in between two thin hexagonal boron nitride (hBN) flakes. The graphene was then side contacted to Cr/Au (3.5/50 nm) metallic electrodes acting as drain and source contacts. In addition, a metal top-gate electrode covering most of the FET channel ($L_{TG}$ = 4.8 µm) was defined on the device, together with a coupled bow tie antenna between top-gate and source electrodes. This antenna ensures an efficient rectification of the incoming THz radiation for a large range of frequencies, via gate-to-source coupling (additional fabrication details are shown in Supporting Information Note 1).

Transport and zero-bias photocurrent measurements in our device were performed in a closed-cycle cryostat, with the chamber temperature varying from 10K up to 300K. We employed two different THz sources to perform the photocurrent experiments. First, a sub-THz source was used to undertake measurements at a frequency of 0.3 THz and then, a quantum cascade laser was used to undertake measurements at frequencies in a range between 2.5 THz up to 4.7 THz (more information about the photocurrent setup can be found in Refs. [21,22]).

## Results and Discussions

First, we measured the transport characteristics of our graphene FET via electrical measurements from 10K up to 300K (see Supporting Information Note 2). We extracted average mobilities, $\mu$, in the device exceeding 70000 cm$^2$V$^{-1}$s$^{-1}$ for both electron and hole carriers at low temperatures (10K). Such values remain high, above 60000 cm$^2$V$^{-1}$s$^{-1}$, even at room temperature (Supporting Information Note 2 contains the measured electrical data as well as details to calculate the carrier mobility). We further estimated the momentum-relaxing scattering time of charge carriers in the device $\tau$ to lie between 0.29 ps and 0.23 ps at 10K and room temperature, respectively. This is calculated with the relation $\tau = m\mu/e$, where $e$ is the elementary charge and $m$ is the effective mass of carriers in single-layer graphene. The latter is given for single layer graphene by $m = \hbar k_F/v_F$, where $v_F$ is the Fermi velocity, $\hbar$ is the reduced Plank constant and $k_F$ the Fermi wave vector ($k_F = \sqrt{\pi n}$, with $n$ as the carrier density).

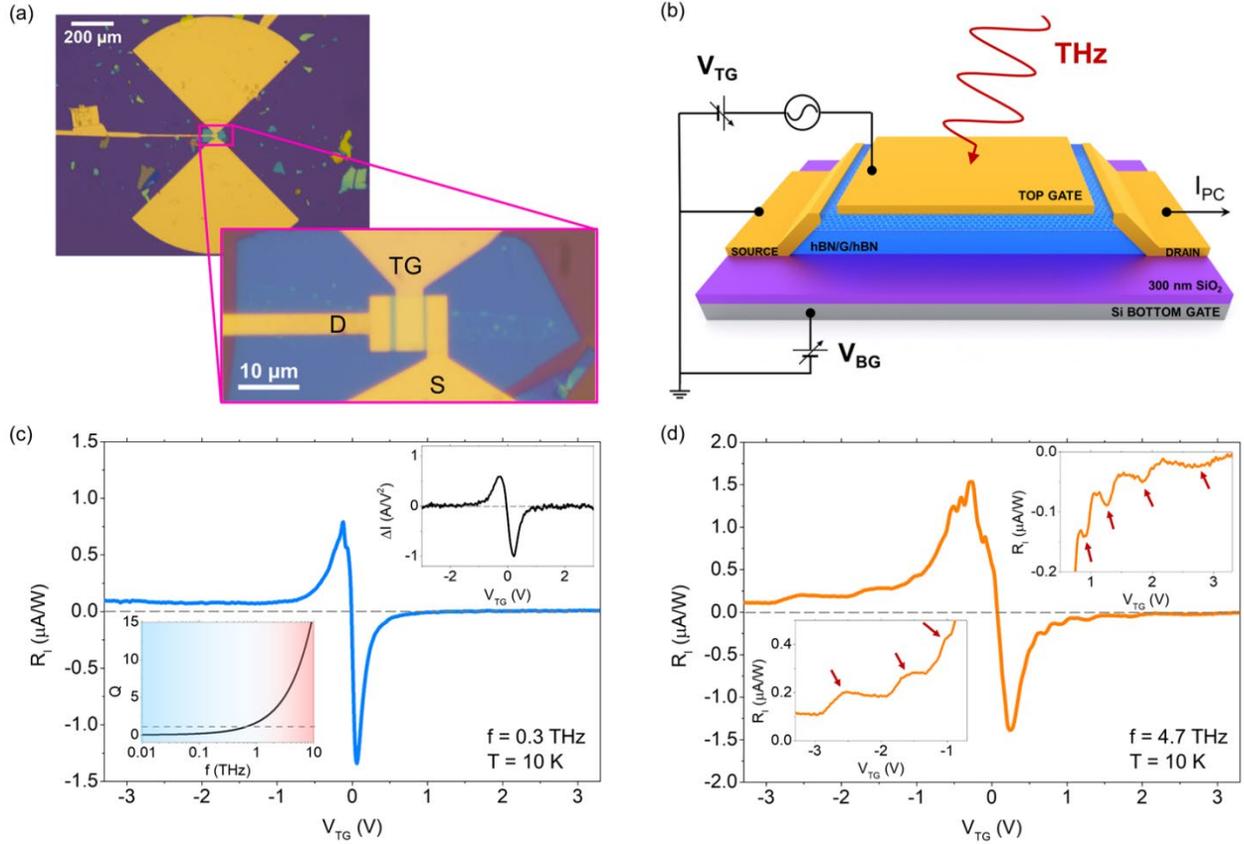

**Figure 1. Graphene-based resonant THz photodetector.** (a) Optical images of the Graphene THz detector with a bow-tie antenna coupled between Source and Top Gate electrodes. The bottom image shows a zoomed view of the device with Source (S), Top Gate (TG) and Drain (D) electrodes labeled. (b) Schematic 3D view of the zero-bias photocurrent measurements of the device. (c) Current responsivity, $R_I$, as a function of the top gate voltage, $V_{TG}$, measured at 0.3 THz. The upper-right inset shows the current responsivity expected from the DC conductivity, following the phenomenological formula[21] $\Delta I = -d\sigma/dV_{TG}$. The bottom-left inset shows the evolution of the Q factor in our device as a function of the excitation frequency. The dashed line corresponds to Q=1. Highlighted areas in blue and red indicate the frequencies ranges in which our device operates the over-damped or weakly damped regimes, respectively. (d) $R_I$ as a function of the top gate voltage measured at 4.7 THz. Inset panels show zoomed in areas of the recorded current responsivity for electron (upper-right) and hole (bottom-left) carriers. Responsivity resonances are highlighted by red arrows in these insets. Here, we note that when the current responsivity is calculated as $R_I = I_{PC} S_T / P S_D$, where $S_T$ is the THz beam spot area and $S_D$ is the detector active area, the device performance reach larger maximum values of ~0.29 A/W at 0.3 THz and ~1.8 mA/W at 4.7 THz.

We carried out zero-bias (i.e. zero source-drain potential, $V_{DS}$) photocurrent measurements at different THz frequencies (Figure 1 (b)). First, we studied the photoresponse of the detector at 10K for an incoming THz frequency of 0.3 THz. The current responsivity of the device, $R_I = I_{PC}/P$, is shown in Figure 1(c), with $P$ being the incoming power of the THz radiation and $I_{PC}$ the measured photocurrent at the drain contact. It is worth noting that, at this radiation frequency, the quality factor value $Q$ is below ~0.5 and thus the photodetector operates in the overdamped regime (see bottom inset of Figure 1 (c) –blue shadowed region). The experimental photoresponse exhibits an antisymmetric shape with respect to the applied top gate potential which flips it sign at the charge neutrality point (CNP). Such trends, together with the appearance of the maxima and minima values of the photocurrent near the CNP and a vanishing photocurrent at large gate voltages, result from the ambipolar charge transport in graphene and agree with previous works reporting non-resonant photodetection in the literature[18–22]. We further highlight that the lineshape of the measured current response w.r.t. the gate potential follows closely the trend predicted by theory[21], $\Delta I = - d\sigma/dV_{TG}$ (see upper inset Figure 1 (c)), with $\Delta I$ being the expected photocurrent and $\sigma$ the DC channel conductivity. The qualitative agreement between both experimental and theoretical curves, with the only sign reversal occurring at the CNP, indicates that the rectified photocurrent in the device is predominantly generated via the so-called plasmonic Dyakonov-Shur (DS) mechanism[17,20]. Minimal discrepancies from the DS theory appear at large negative top gate voltages, where the experimental current responsivity shows a rather small responsivity offset ≈ 0.1 μA/W (an order of magnitude lower than the maximum $R_I$ measured), instead of the zero-photocurrent value expected from a pure plasmonic DS mechanism at large bias conditions. Such behaviour may result from an additional rectification effect occurring due to the presence of pn juntions at the metal-graphene contact[20,21].

Next, we measure the photoresponse of the device at 10K but a higher frequency, 4.7 THz (Figure 1 (d)). The quality factor at this radiation frequency is characterized by $Q \gg 1$ ($Q \sim 8.6$), and thus the device operates in the resonant (weakly-damped) regime (see Figure 1 (c), bottom inset). Intriguingly, the current photoresponse recorded at this higher frequency exhibits not only the characteristic antisymmetric line shape with respect to the applied gate voltage (similar to the broadband case depicted in Figure 1 (c)) but also marked oscillations on both electron and hole sides emerged (see arrows in top-right and bottom-left insets in Figure 1(d), respectively). Such oscillations, which are dependent on the carrier density, constitute the hallmark of resonant operation in a FET photodetector[17,24–28]. They are the result of plasmon resonances occurring in the graphene channel because of the reflection of the plasma waves at the end of the channel and the interference of both reflected and incoming waves. Under such conditions, the graphene device acts like a Fabry-Perot resonant cavity for the propagating graphene plasmons under external THz excitation. The multiple ridges presented in $R_I$ are the result of the crossover from destructive to constructive interferences of the incoming and reflected waves. Subsequently, peaks represent waves with a number of oscillation modes which are by one higher or lower than the neighboring crests. The mode number is tunable with both the length of the top gate ($L_{TG}$) and the density of the charge carriers (controlled via the applied gate voltage, $V_{TG}$) in the system[28]. Importantly, the intensity of such resonances strongly depends on the plasmonic cavity length ($L_{TG}$) and the plasmon propagation length ($L_P$, which is larger than the $1/e$-decay length $L_d = s\tau$ of the plasma wave and depends on the signal-to-noise ratio at which small modulations of $R_I$ can still be detected), leading to two different scenarios[17]. When $L_{TG} \lessapprox L_P$, propagating plasmons can reach the end of the channel before a total decay, creating interferences between the incoming and reflected waves at least at the end of the channel, if not along its total length (see Supporting Movie 1). Such case gives rise to different characteristic resonant modes as a function of the carrier density or the incoming frequency. Then, if $L_{TG} \gg L_P$, propagating plasmons in the system decay before reaching the end of the cavity (see Supporting Movie 2), giving rise a rectified photocurrent indistinguishable to the one expected in the non-resonant scenario[17].

For completeness, we additionally measured the photoresponse at different frequencies (range 2.5 THz – 4.7 THz), all within the resonant regime or weakly damped scenario ($Q \gg 1$). Figure 2 (a) highlights the evolution of the emerged photoresponse oscillations within this frequency range. For simplicity and clarity, we present the normalized current responsivity, $R_I^N$, with respect to the photocurrent maximum observed close to the CNP. Interestingly, the current photoresponse as a function of the gate voltage exhibits oscillations at all these measured frequencies, but the visible number of oscillations strongly depends on the THz frequency. In particular, the number of peaks decreases when lowering the excitation frequency. Further frequency-dependent measurements can be found in Supporting Information Note 3.

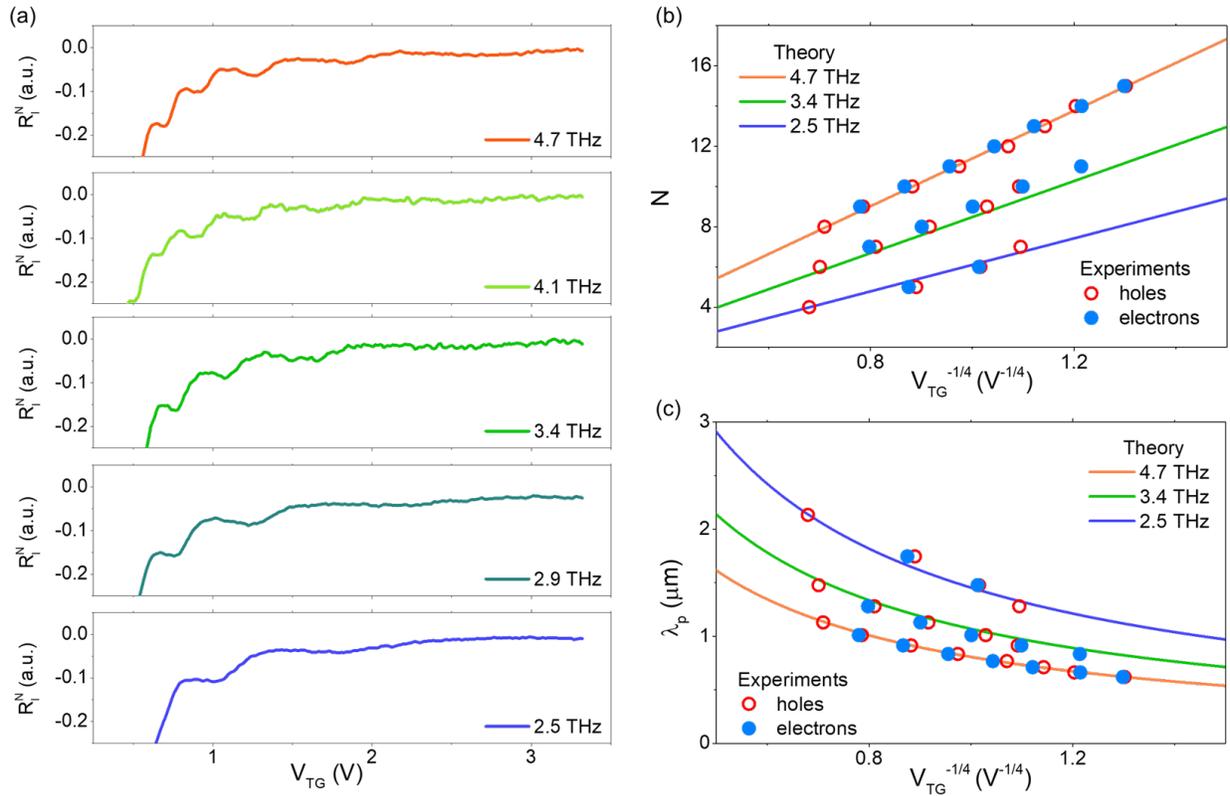

**Figure 2. Frequency dependence of resonant THz photodetection.** (a) Normalized current responsivity, $R_I^N$, as a function of the top gate voltage at the electron side for different frequencies in the range 2.5 THz - 4.7 THz ($Q \gg 1$ in the device for all these frequencies). All measurements in these five panels were performed at 10K. In this panel, the measured current responsivity is normalized ($R_I^N$) with respect to the photocurrent maximum recorded close to the CNP for an easier comparison of all recorded data at the different frequencies. (b) Resonant mode number, $N$, of the local minima in the $R_I^N$ curves and (c) corresponding plasmon wavelength, $\lambda_p$, as a function of $V_{TG}^{-1/4}$ for three selected frequencies from panel (a). Solid lines correspond to the calculated theoretical dependence following equation (4) and symbols represents the extracted values from experiments.

The observed oscillations of the current photoresponse when sweeping $V_{TG}$ can be further analyzed in the following way. Excited plasmons in gated two-dimensional systems follow the linear dispersion law, $\omega = sk$, where $s$ is the plasma wave velocity and $k$ is the real part of the angular wave number[24,25]. Plasma wave velocity is defined as:

$$s = \sqrt{\frac{e}{m}|V_{TG}|} \qquad (1)$$

And resonances should emerge when the real part of the wave number is given by:

$$k = \frac{\pi}{2L_{TG}}(2N+1), \quad N = 0, 1, 2 \ldots \qquad (2)$$

Importantly, the effective mass, $m$, in single-layer graphene[18,31] is dependent on the applied gate voltage as $m = \frac{\hbar k_F}{v_F} = \frac{\hbar}{v_F}\sqrt{\frac{\pi C_{ox}|V_{TG}|}{e}}$ (in the former expression, $C_{ox}$ is the thin-oxide gate capacitance per unit area and $v_F$ is the Fermi velocity of the charge carriers). Thus, by replacing $m$ into equation (1), the plasma wave velocity in MLG can be rewritten as:

$$s = \sqrt{\frac{e\, v_F}{\hbar}\left(\frac{e|V_{TG}|}{\pi C_{ox}}\right)^{\frac{1}{4}}} \qquad (3)$$

Then, using the plasmon dispersion law with equations (2) and (3), one can easily deduce the relation between the resonant mode number, $N$, and the applied gate voltage, $V_{TG}$, for single-layer graphene:

$$N = \frac{L_{TG}\omega}{\pi\sqrt{\frac{e\, v_F}{\hbar}\left(\frac{e|V_{TG}|}{\pi C_{ox}}\right)^{\frac{1}{4}}}} - \frac{1}{2} = \alpha\omega|V_{TG}|^{-1/4} - \frac{1}{2} \qquad (4)$$

Following equation (4), $N$ is expected to have a linear dependence with $\omega$ and $V_{TG}^{-1/4}$. We verified that the experimental resonant peaks appearing at the different measured frequencies in our device (Figure 2 (a)) follow the predicted $V_{TG}^{-1/4}$ dependence. In particular, Figure 2(b) shows the extraordinary agreement between the calculated theoretical dependence $N(\omega, V_{TG})$ given by equation (4), and the values of $N$ extracted from the experimental data. We note that, in comparison with systems with parabolic bands[24–26], graphene's linear energy-momentum results in a notably distinct dependence of $N$ with the applied voltage (systems with parabolic energy bands exhibit a relation dependence of $N \propto V_{TG}^{-1/2}$ instead[28]). In consequence, the first resonant modes in our single-layer graphene THz detector ($N < 6$) are not accessible in the recorded $V_{TG}$ range at the highest measured frequency 4.7 THz due to the $V_{TG}^{-1/4}$ dependence of $N$ introduced in equation (4). For instance, at 4.7 THz, the resonant mode $N = 2$ is expected to occur for gate potentials larger than 250 V, values which are not reachable in common experimental devices. Resonant modes below 6 are only experimentally accessible in our device when decreasing the excitation frequency down to 2.5 THz (see Figure 2 (b)).

The observed resonant modes in the THz photoresponse can be further utilized to extract significant information of the propagating graphene plasmons[23,28,32]. Such information includes the plasmon lifetime ($\tau_p$) and the plasmon wavelength ($\lambda_p$). We calculated the plasmon lifetime by using the width of the characteristic resonant peaks at the half-height and the gate voltage at which plasmon resonances arise (see Supporting Information Note 4). The resulting values for $\tau_p$ were found to be around 0.6 ps, which are larger than the aforementioned scattering time values extracted from transport analysis. Similarly, the plasmon wavelength can be determined from the measured resonances observed in the photocurrent with respect to the gate voltage and excitation frequency, following the relation[23,28,32] $\omega = 2\pi s/\lambda_p$. The obtained values for $\lambda_p$ range between 600 nm and 2.1 μm (see Figure 2 (c)) for the studied range of THz frequencies (2.5 – 4.7 THz).

These plasmon wavelength values lead to compression ratios ($\lambda_o/\lambda_p$, with $\lambda_o = c/f$ being the wavelength of the incoming THz radiation in free-space) as high as 110 (see details in Supplementary Note 5). The ratio agrees well with the extreme light compression and nano-scale confinement occurring in graphene devices at THz frequencies reported in previous works[28,32,33].

Finally, we measured the evolution of the plasmonic resonances at 4.7 THz when raising up the temperature, *T*. Figure 3 (a) shows the zero-bias photoresponse as a function of the top gate potential for the hole-side (negative $V_{TG}$) and the electron-side (positive $V_{TG}$) at four selected temperatures within the range 10K-300K. Importantly, the observed resonant peaks and dips persist up to 300K both for electron and hole conduction (see also additional data in Supplementary Note 6). Thus, these measurements demonstrate the resonant detection of THz radiation at room temperature in non-biased FET devices made from single-layer graphene.

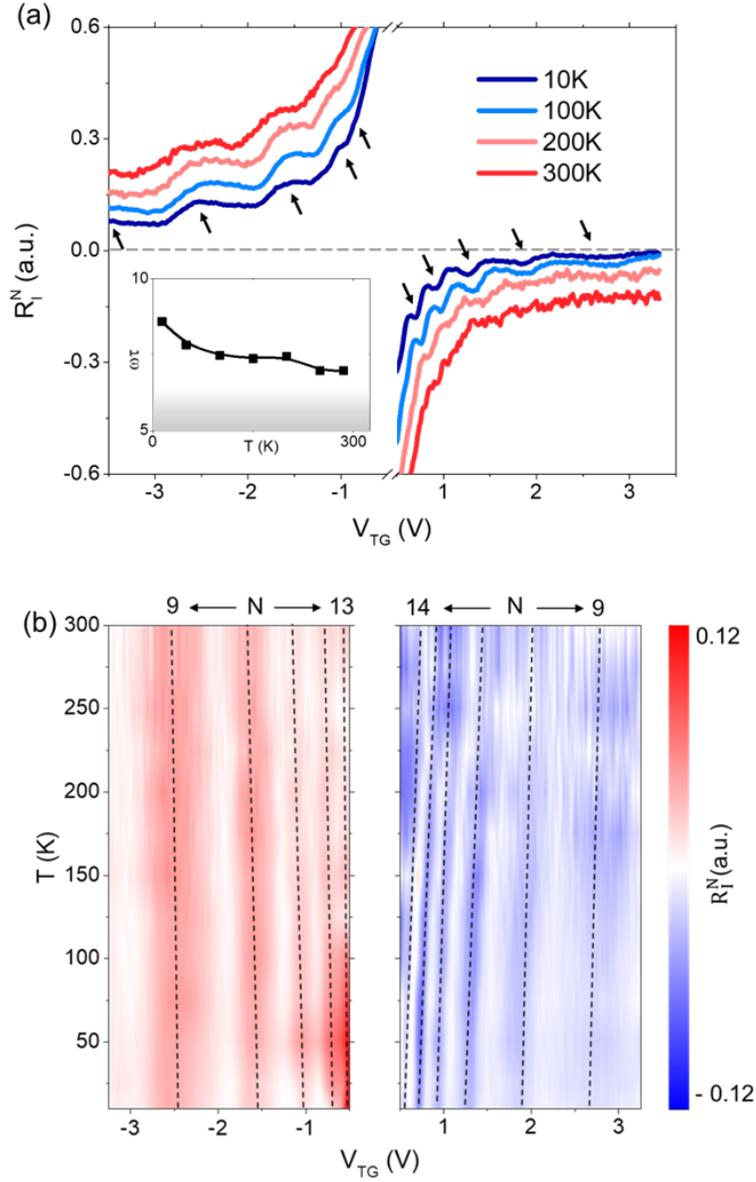

**Figure 3. Temperature evolution of the plasmonic resonances.** Zero-bias normalized photocurrent as a function of the top gate voltage at four selected temperatures from 10 K up to 300K (room temperature) at both the hole and electron regions for an incident radiation of 4.7 THz. For an easier visualization, the temperature-dependent photoresponses shown in the panel are normalized with respect to the maximum near the CNP (as done in Figure 2(a)) and the curves are vertically shifted. The bottom inset shows the temperature evolution of the quality factor $Q=\omega\tau$ (b) Color mapping of the normalized responsivity, $R_I^N$ (after subtraction of the non-resonant back-ground) as a function of the top gate voltage at 4.7 THz for all measured temperatures. Vertical dashed lines highlight the evolution of the different observed resonant modes with temperature.

The presence of the photoresponse oscillations with respect to the $V_{TG}$ and their detailed evolution with temperature are clearly visible in Figure 3 (b). In particular, this panel presents the measured photoresponse when excluding (i.e. subtracting) the broadband contribution for all measured temperatures within the range of 10K- 300K. We stress the fact that resonant peaks and dips appear in the map approximately at the same carrier density (i.e. same value of $V_{TG}$) for all temperatures. This observation agrees well with equation (4), which does not contain any explicit dependence of the position of the resonances on *T*. We notice that gate-tunable photoresponse resonances shown in Figure 3 (a) are more evident at the hole side (negative gate voltages) than the electron side (positive gate voltages). This is also seen in Figure 3 (b) when the resonances are displayed as a function of the temperature. We argue that this could be caused by slightly larger mobilities on the hole side w.r.t. the electron side in our devices (see Supporting Information Note 2). Moreover, the amplitude of the photocurrent oscillations measured at room temperature in electron or hole conduction depends ultimately on the device (see Supporting Information Note 7 showing stronger and more evident resonances measured at room temperature in a second photodetector).

The fact that devices made from high-quality, single-layer graphene exhibit clear and unambiguous evidences of resonant responsivity at room temperature (including the appearance of oscillations of the zero-bias photoresponse with respect to the gate voltage – or equivalently carrier density) is extremely relevant for applications. Up to date, robust signatures of this resonant regime had only been reported to occur at cryogenic temperatures in semiconductors such as bilayer graphene[28] or 2D electron gases made of III-V materials [24–26].
Only some experimental indications have been interpreted as arising from resonant detection in III-V field-effect transistors operating at room temperature[34,35]; but these are vague and rely on the application of a large source-to-drain bias (the application of a source-to-drain dc voltage or current shifts the system towards a resonant regime[36], but also increase the noise of the rectified signal). In contrast, our study (Figure 3) shows strong and univocal plasmonic resonant oscillations in zero-biased photocurrent measurements performed at room temperature.

We argue that the robust observation of the resonant regime in high-quality single-crystal graphene results from the large room-temperature mobility of the charge carriers in this material[29,30], which for our devices is larger than 60000 cm$^2$V$^{-1}$s$^{-1}$ (See Supplementary Note 2). Such a value leads to a transport scattering time $\tau = 0.2$ ps even at room temperature and to a quality factor Q>>1 (Q > 6) at an excitation frequency of 4.7 THz (bottom inset of Figure 3 (a) shows the evolution of Q with temperature in the device). Since the condition Q>>1 is fulfilled, micrometer-size devices made of high-quality monolayer graphene can robustly operate in the weakly damped regime at room temperature and show resonant detection. In contrast, other semiconductor materials have intrinsic carrier mobilities which are around or even below 5000 cm$^2$V$^{-1}$s$^{-1}$ at room temperature[37], which impedes the observation of resonant detection at room temperature. This is even the case of bilayer graphene[28], system which also has a lower intrinsic room-temperature mobility values (~15000 cm$^2$V$^{-1}$s$^{-1}$) than monolayer graphene due to the presence of additional intrinsic scattering sources including shear phonon scattering[38] or significantly larger electron-hole collisions[39].

## Conclusions

In summary, we have studied the zero-bias photoresponse of a high-mobility monolayer graphene FET subjected to THz radiation. The operation of the device is perfectly tuned between non-resonant and resonant regime depending on the frequency of the incoming radiation. In particular, the resonant regime is univocally demonstrated by the measured oscillations present in the gate-voltage-dependent photocurrent. These oscillations are dependent on both the carrier density in the channel and the frequency of the THz radiation. We demonstrate that such univocal fingerprints of resonant THz photodetection are not only visible at cryogenic temperatures but also at room temperature. To the best of our knowledge, this is the first time that resonant THz photodetection has been robustly observed at room temperature without application of a large drain current bias (which is undesirable for a proper detector operation).

From an application point of view, these findings pave the way for the design and development of a new generation of (graphene-based) plasmonic resonant photodetectors operating at room temperature. The application space of such systems is significant in the THz and mid-infrared regime[6,7] allowing the realization of emerging and potential technologies at these relatively unexploited but relevant frequencies, including modulators, filters, polarizers, emitters and selective photodetectors, among many others; as well as the confinement and manipulation of the electromagnetic fields below the classical diffraction limit[32].

## Acknowledgments

Authors thank the support from the Ministry of Science and Innovation (MCIN) and the Spanish State Research Agency (AEI) under grants (PID2021-126483OB-I00, PID2021-128154NA-I00) funded by MCIN/AEI/10.13039/501100011033 and by "ERDF A way of making Europe". This work has been also supported by Junta de Castilla y León co-funded by FEDER under the Research Grant numbers SA103P23. J.M.C acknowledges financial support by the MCIN and AEI "Ramón y Cajal" program (RYC2019-028443-I) funded by MCIN/AEI/10.13039/501100011033 and by "ESF Investing in Your Future". J.M.C also acknowledges financial of the European Research Council (ERC) under Starting grant ID 101039754, CHIROTRONICS, funded by the European Union. Views and opinions expressed are however those of the author(s) only and do not necessarily reflect those of the European Union or the European Research Council. Neither the European Union nor the granting authority can be held responsible for them. K.W. and T.T. acknowledge support from the JSPS KAKENHI (Grant Numbers 21H05233 and 23H02052) and World Premier International Research Center Initiative (WPI), MEXT, Japan. The work in Frankfurt is supported by DFG projects RO 770/40-2 and RO 770/53-1. J.A.D-N thanks the support from the Universidad de Salamanca for the María Zambrano postdoctoral grant funded by the Next Generation EU Funding for the Requalification of the Spanish University System 2021–23, Spanish Ministry of Universities. Authors also acknowledge USAL-NANOLAB for the use of Clean Room facilities.

# Supplementary Information

## Supplementary Note 1. Fabrication details

Monolayer graphene and hexagonal-boron-nitride (hBN) were mechanically exfoliated on a Si substrate with 300 nm of $SiO_2$ thermally growth and identified via optical contrast using an optical microscope. The exfoliated hBN flakes were also characterized by a Stylus Profilometer (Bruker DektakXT®) to estimate their thickness, obtaining values of 28 nm for the top and 32 for the bottom. The graphene-based heterostructure was fabricated by using a state-of-the-art dry-stacking technique[1,2] to encapsulate the monolayer graphene in between two thin hBN flakes (See Figure S1). Importantly, we have intentionally selected a large and straight graphene flake to naturally define the width, W, of device channel (W ≈ 4.5 μm). Initially, the sample was patterned via electron beam lithography (EBL) and dry etching in a $SF_6$ atmosphere (20 ºC, 40 sccm and 75W) to remove unwanted flakes transferred near the graphene-based stack. Then, Drain and Source side metallic contacts and the bowtie antenna were fabricated via EBL, dry etching in a $SF_6$ atmosphere (10 ºC, 40 sccm and 75W) and e-beam evaporation of 3.5 nm Cr and 55 nm Au. Here one of the bows of the antenna was connected to the Source contact. Finally, another round of EBL and e-beam evaporation of 5 nm Cr and 40 nm Au was undertaken to fabricate the Top Gate and connect it to the other bow of the antenna.

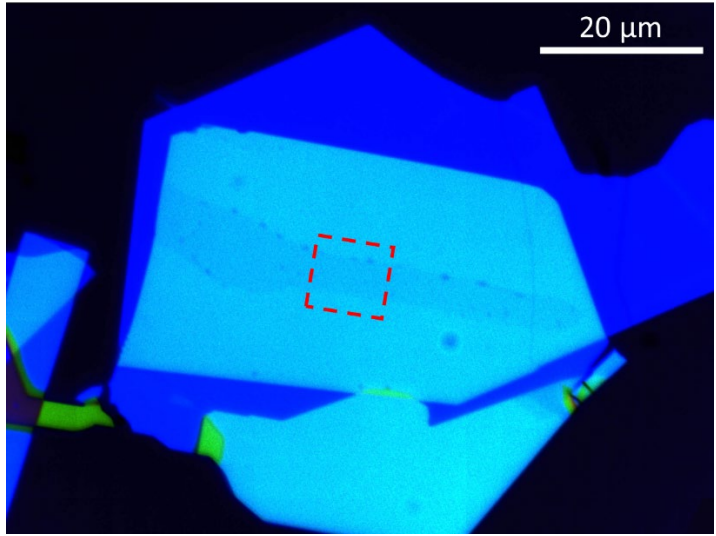

**Figure S1.** Optical image of the hBN/Monolayer Graphene/hBN stack. Red-dashed line highlights the area selected for the fabrication of the Terahertz detector.

# Supplementary Note 2. Transport measurements

Two-terminal dc transport measurements were performed via lock-in technique where a pseudo-dc current (10 nA at 11.33 Hz) was injected in the device and the generated voltage drop across source and drain electrodes was recorded by a lock-in amplifier (SR860). The applied top gate potential was generated with a dc voltage generator (Keithley 2614B). Figure S2 (a) shows the two-terminal channel resistance, $r_{ch}$, w.r.t the top gate potential, $V_{TG}$, measured at several, equally spaced temperatures in the range 10K-300K. Instead, Figure S2 (b) shows the channel resistance curves at four selected temperatures. The measured resistance exhibits a bell shape characteristic of graphene devices[2–5], with the resistance maxima occurring at the charge neutrality point (CNP) and a decrease in magnitude when applying a top gate potential away of the CNP.

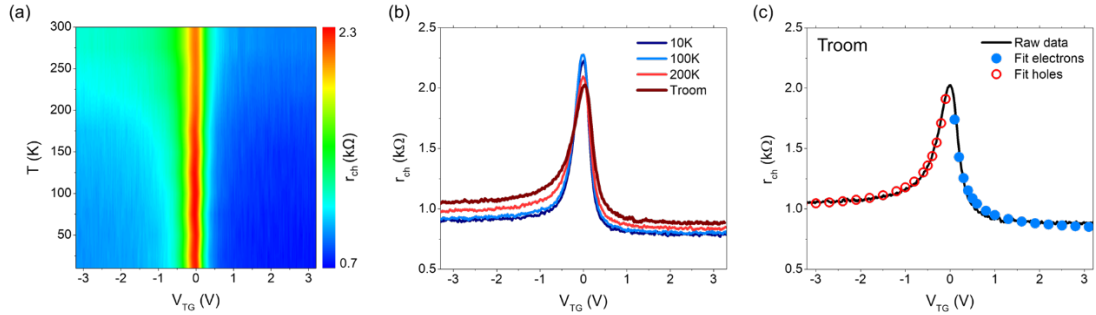

**Figure S2.** (a) Two-terminal resistance, $r_{ch}$, w.r.t. the top gate voltage measured at several temperatures. (b) Two-terminal resistance w.r.t. the top gate potential at four selected temperatures extracted from panel (a). (c) Experimental (black line) and fitted (blue and red dots) channel resistance data at room temperature

In order to estimate average values of carrier mobility in our device, we have fitted the two-terminal resistance curves following the model presented by L. Gammelgaard and co-authors[6] characterized by the equation:

$$r_{ch} = R_c + \frac{L/W}{\sqrt{\left(\frac{L/W}{(R_{CPN} - R_c)e\mu}\right)^2 + n^2 e\mu}} \tag{1}$$

where $R_c$ is the parasitic resistance (i.e. the two contact resistances and the resistance of the ungated regions in the graphene channel), $L$ and $W$ are the channel length and width respectively, $R_{CNP}$ is the resistance value at the CNP, $e$ is the elementary charge, $\mu$ is the carrier mobility and $n$ is the carrier concentration approximated by $n = C_{ox}V_{TG}^*/e$ where $C_{ox}$ is the gate oxide capacitance per unit area and $V_{TG}^*$ is the applied top gate voltage with respect to the CNP. We have obtained values of average carrier mobilities exceeding 70000 cm$^2$V$^{-1}$s$^{-1}$ at temperatures around 10 K and remain as high as 60000 cm$^2$V$^{-1}$s$^{-1}$ at room temperature.

To underpin the high-mobility values of charge carriers in our samples, we have additionally fabricated a multi-terminal Hall bar device using the same procedure as described in the manuscript and Supplementary Note 1. The longitudinal device resistance was measured via four-terminal configuration at different back-gate voltages (see Figure S3). The field-effect mobility μ was calculated using the Drude model of conductivity given by the formula $\sigma = \mu_i e n_i$, where $\sigma$ is the is the electrical conductivity, $e$ is the elementary charge and $n_i$ is the electron or hole carrier densities. We observed maximum carrier mobilities exceeding 150000 cm$^2$/Vs for holes and electrons in the samples at 10K (see Figure S3).

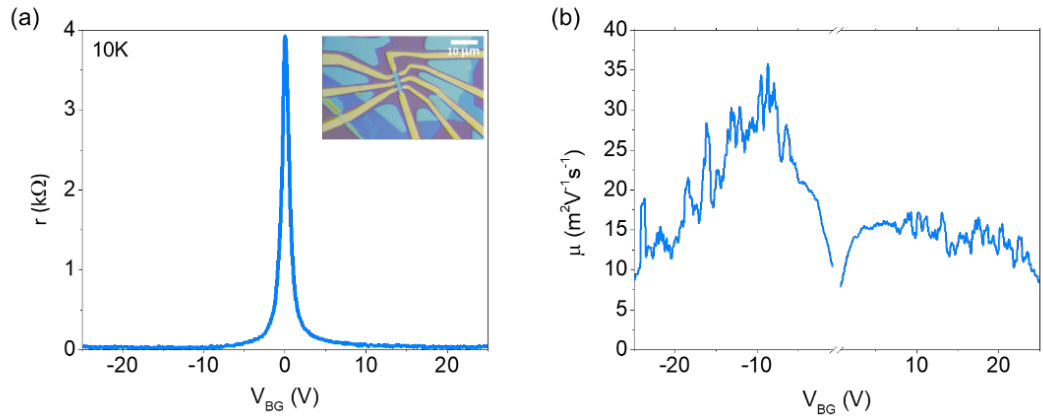

**Figure S3.** (a) 4-terminal resistance, r, as a function of the back-gate voltage, $V_{BG}$. Inset panel shows the optical photograph of the Hall bar device. (b) Carrier mobility as a function of the back-gate potential. Temperature was fixed at 10K.

# Supplementary Note 3. Additional frequency dependence data

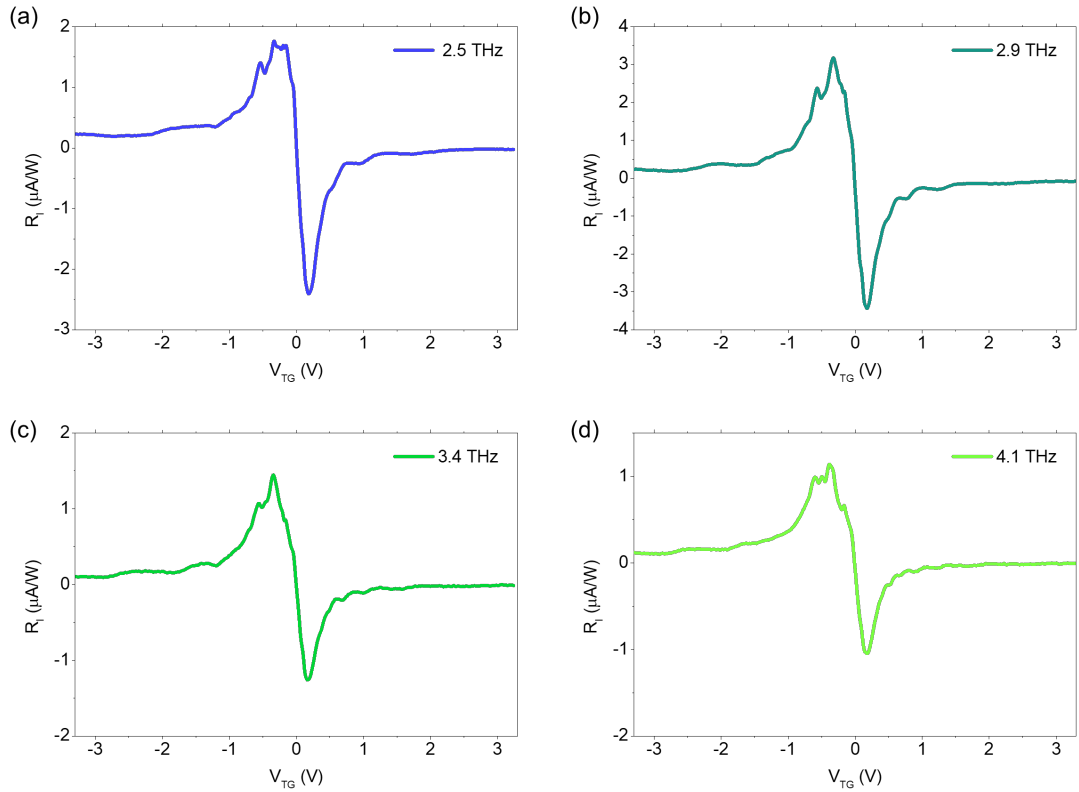

**Figure S4.** Current responsivity, $R_I$, as a function of the top gate voltage, $V_{TG}$, measured for an incoming radiation of (a) 2.5 THz, (b) 2.9 THz, (c) 3.4 THz and (d) 4.1 THz. Temperature was fixed at 10K in all the measurements.

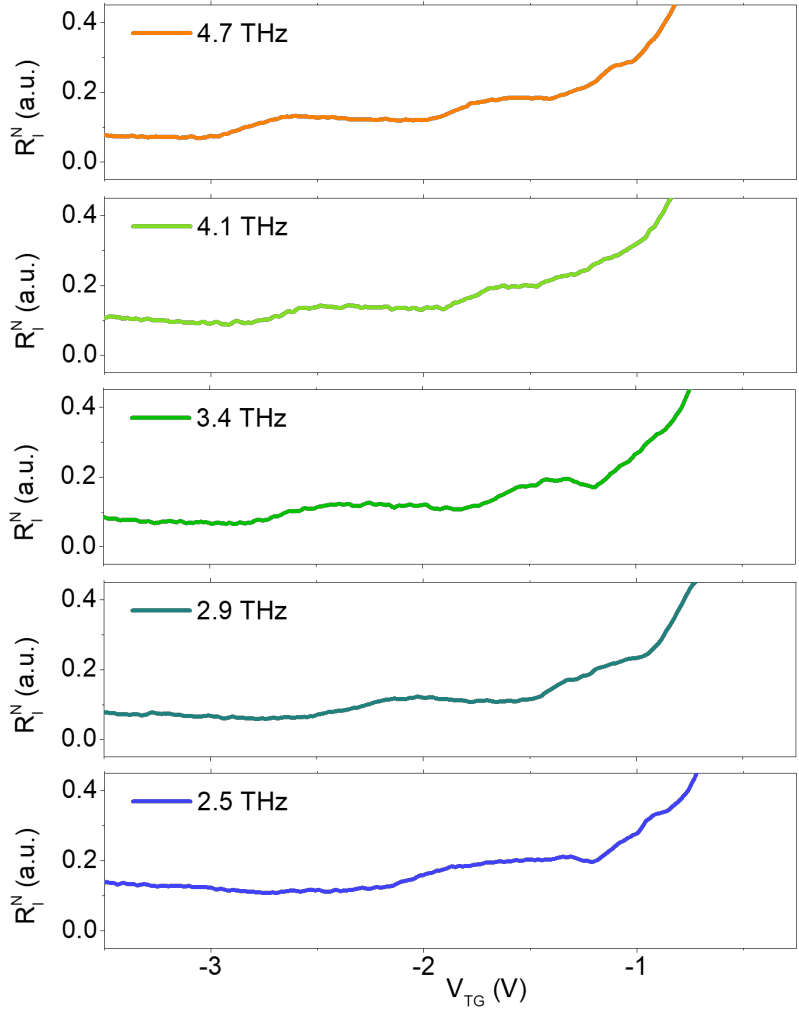

**Figure S5.** Zoomed normalized current responsivity, $R_I^N$, as a function of the top gate voltage, $V_{TG}$, at the hole side for a set of incoming frequencies in the resonant regime (2.5 THz - 4.7 THz, all frequencies making $Q \gg 1$). All measurements in these five panels were performed at 10K.

# Supplementary Note 4. Graphene plasmons in single-layer graphene FET devices

As stated in the main text, the resonant mode number in monolayer graphene with the applied gate voltage follows the formula:

$$N = \frac{L_G \omega}{\pi \sqrt{\frac{e v_F}{\hbar}} \left(\frac{e}{\pi C_{ox}}\right)^{\frac{1}{4}}} |V_G|^{-\frac{1}{4}} - \frac{1}{2} \qquad (2)$$

Which predict a linear dependence of the different resonant modes accessible in our plasmonic cavity with respect to $|V_G|^{-1/4}$, the frequency ω and the cavity length, $L_G$.

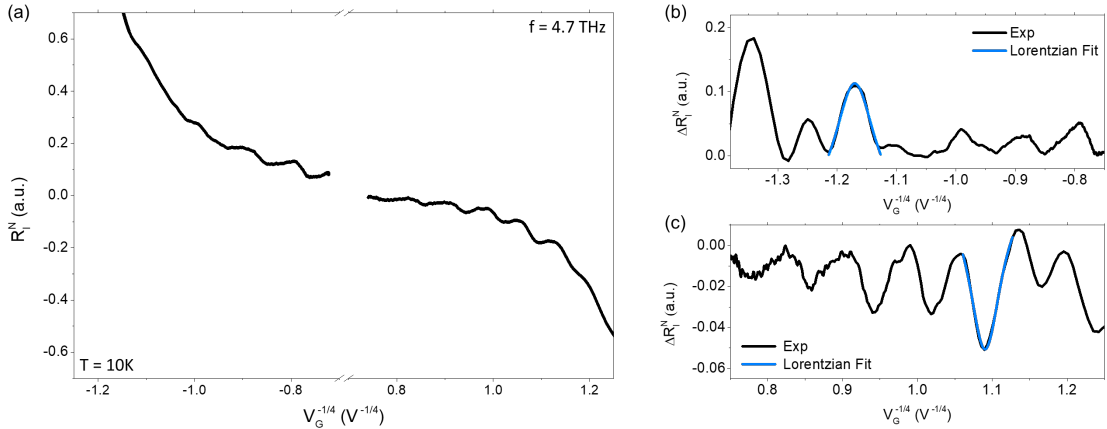

**Figure S6.** (a) Normalized current responsivity, $R_I^N$, as a function of $V_G^{-1/4}$ for an excitation frequency of 4.7 THz at 10K. (b) and (c) Experimental normalized current responsivity (black line) upon removing the non-resonant background as a function of $V_G^{-1/4}$ for the hole side (b) and electron side (c). Solid blue lines correspond to the Lorentzian fitting with a FWHM = 0.056 $V^{-1/4}$ for the electron side and FWHM = 0.085 $V^{-1/4}$ for the hole side which corresponds to τ = 0.6 ps and τ = 0.47 ps respectively for an incoming 4.7 THz.

First, we have analyzed the tunability of the different resonant modes with $|V_G|^{-1/4}$ for an incoming THz radiation of 4.7 THz at 10K. Figure S5 (a) shows the normalized current responsivity as a function of $|V_G|^{-1/4}$. Figure S5(b) and (c) show a similar graph where, for clarity, the non-resonant signal back-ground in the current responsivity has been removed. As expected, the emerged oscillations (corresponding to the different resonant modes appearing at the hole and electron sides) are located equidistantly with respect to the increase of $|V_G|^{-1/4}$ in clear agreement with equation (2). Moreover, we obtain the plasmon lifetime, $\tau_p$ from panels (b) and (c). Adapting the theory developed by Bandurin and coworkers for the case of bilayer graphene[7], $\tau_p$ in the case of single-crystal graphene can be extracted from the peak-width at the half-heigh according to:

$$\frac{FWHM}{V_G^{-1/4}} = \frac{1}{\omega\tau} \qquad (3)$$

By using a Lorentzian fit in the current responsivity curves (see Figure S5 (b)) $\tau_p$ were found to be around 0.6 ps for the electron side and 0.47 ps for the hole side. These values are of the same order of magnitude than the momentum relaxation time extracted from transport analysis but slightly larger ($\tau \approx 0.29$ ps).

Moreover, we have performed simulations analysis of the dependence of the plasmonic cavity length to observe the propagation of plasma waves in time (See supporting movies 1 and 2). When the plasmonic cavity length is much larger than the plasmon propagation length (i.e. L_G >> L_P), propagating plasmons in graphene decay before reach the end of the cavity (See supplementary movie 2). In this scenario, the rectified photocurrent do not differ for both resonant (Q >> 1) or non-resonant regime (Q << 1, case of our device at the measured frequency 0.3THz), and hence no frequency and gate dependent oscillations appear in the photoresponse. When $L_G \approx L_P$, (case of our device at frequencies > 2.5THz) propagating plasma waves slightly decay and they can reach the end of the channel, be reflected and therefore creating a constructive and destructive interferences in the channel in a quasi-standing wave scenario. In this regime, a notable dependence of the photoresponse on the incoming THz frequency and the applied gate voltage is observed (See supplementary movie 2).

# Supplementary Note 5. THz field confinement in FET devices made of single-layer Graphene

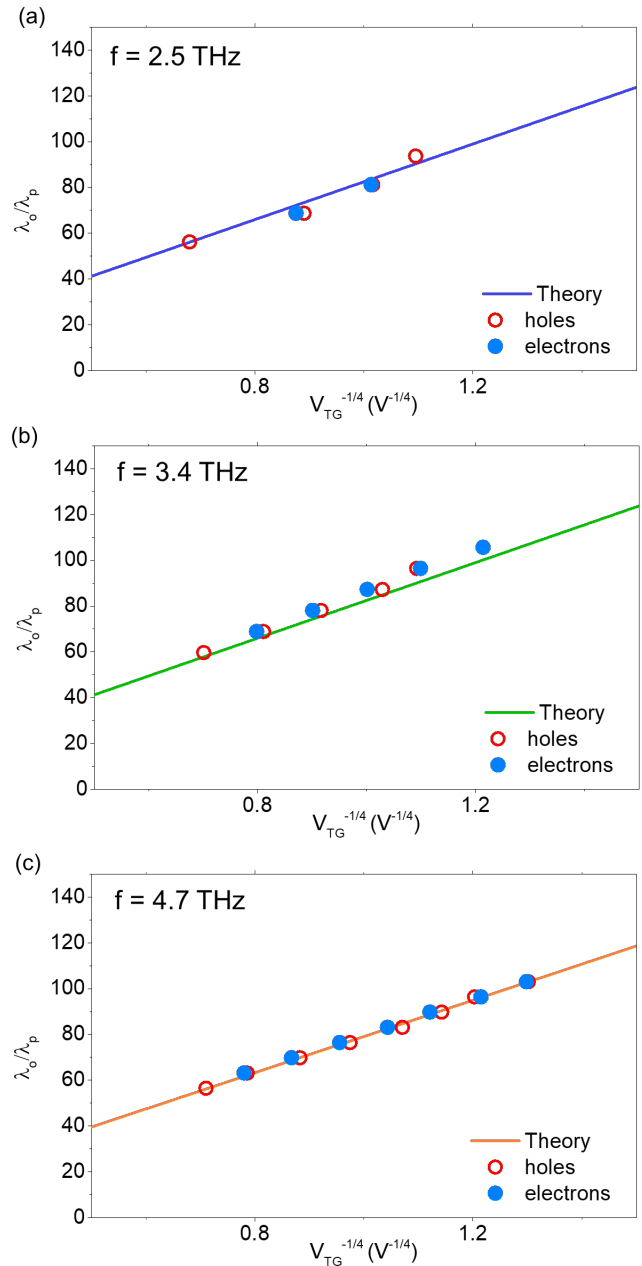

**Figure S7.** Compression ratio, $\lambda_o/\lambda_p$, as a function of $V_{TG}^{-1/4}$ for excitation frequencies of 2.5 THz (a), 3.4 THz (b) and 4.7 THz (c) at 10K.

# Supplementary Note 6. Room temperature resonant THz photodetection

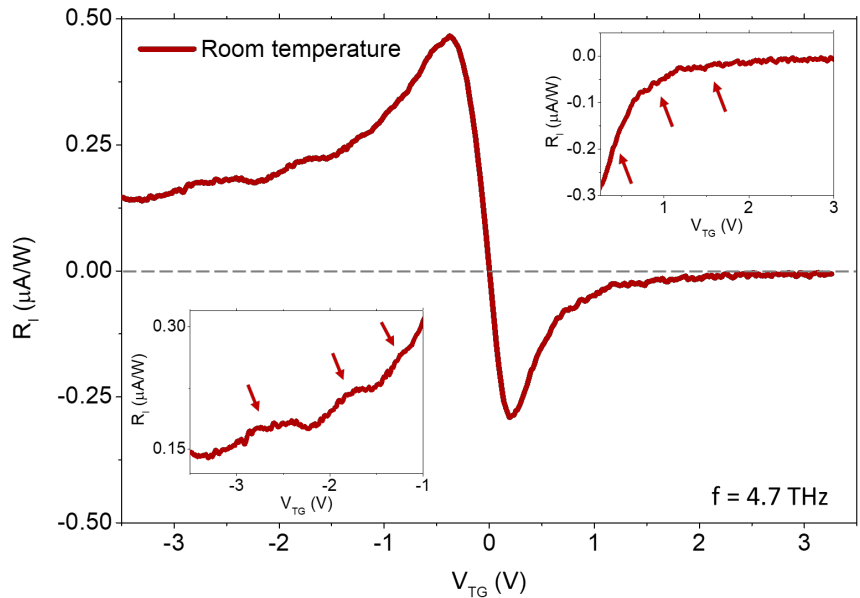

**Figure S8.** Current responsivity, $R_I$, as a function of the top gate voltage, $V_{TG}$, measured at 4.7 THz at room temperature. Inset panels show a zoomed area of the recorded current responsivity for electron (upper-right) and hole (bottom-left) carriers where resonant peaks are highlighted by red arrows.

# Supplementary Note 7. Additional THz photodetectors

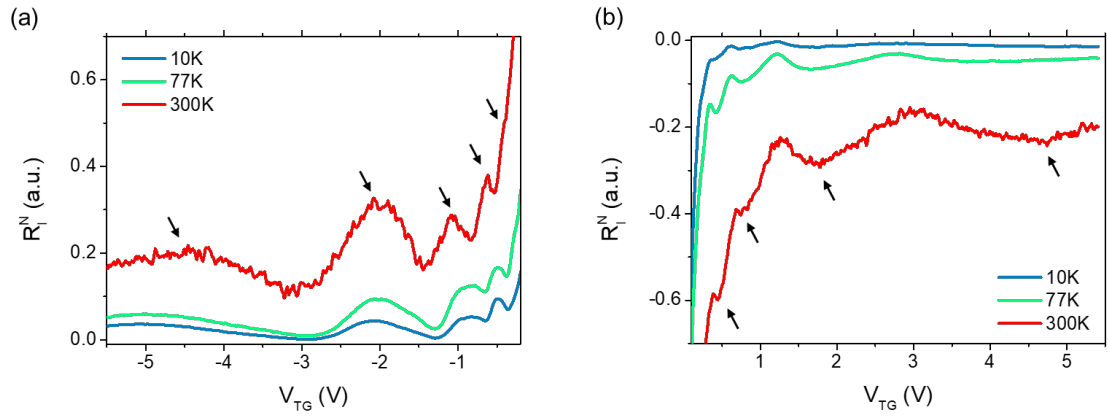

**Figure S9. Additional device showing room-temperature resonant photoresponse.** Zero-bias normalized photocurrent as a function of the top gate voltage for three selected temperatures of 10 K, 77K and 300K (room temperature) at the hole (a) and electron (b) regions for an incident radiation of 2.5 THz for a second device with L = 4 μm. For an easier visualization, the temperature dependence photoresponses shown in the panel are normalized with respect to the maxima near the CNP.